\documentclass[conference,letterpaper]{IEEEtran}
\usepackage{amsthm,amssymb,color,cite}
\usepackage[utf8]{inputenc} 
\usepackage[T1]{fontenc}
\usepackage{url}
\usepackage{ifthen}
\usepackage{cite}
\usepackage[cmex10]{amsmath} 

\usepackage{xcolor}
\usepackage{textgreek} 
\usepackage[normalem]{ulem}
\DeclareMathAlphabet{\mathbb}{U}{bbold}{m}{n}

\newtheorem{theorem}{Theorem}

\newtheorem{lemma}{Lemma}

\theoremstyle{definition}
\newtheorem{remark}{Remark}

\newcommand{\F}{\mathbb{F}}

\usepackage{mathtools}

\let\set\relax
\DeclarePairedDelimiter\set{\{}{\}}

\allowdisplaybreaks

\begin{document}
\title{On the Worst-Case Complexity of Gibbs Decoding for Reed--Muller Codes} 


\author{%
  \IEEEauthorblockN{Xuzhe Xia}
  \IEEEauthorblockA{University of British Columbia\\
                    Vancouver, Canada\\
                    xia2019@student.ubc.ca}
  \and
  \IEEEauthorblockN{Nicholas Kwan}
  \IEEEauthorblockA{University of Toronto\\
                    Toronto, Canada\\
                    nick.kwan@mail.utoronto.ca}
\and
\IEEEauthorblockN{Lele Wang}
  \IEEEauthorblockA{
                    University of British Columbia\\
                    Vancouver, Canada\\
                 lelewang@ece.ubc.edu}
}

\maketitle



\begin{abstract}
    Reed--Muller (RM) codes are known to achieve capacity on binary symmetric channels (BSC) under the Maximum a Posteriori (MAP) decoder. However, it remains an open problem to design a capacity achieving polynomial-time RM decoder. Due to a lemma by Liu, Cuff, and Verd\'u, it can be shown that decoding by sampling from the posterior distribution is also capacity-achieving for RM codes over BSC. The Gibbs decoder is one such Markov Chain Monte Carlo (MCMC) based method, which samples from the posterior distribution by 
    flipping message bits according to the posterior, and can be modified to give other MCMC decoding methods.
    In this paper, we analyze the mixing time of the Gibbs decoder for RM codes.
    Our analysis reveals that the Gibbs decoder can exhibit slow mixing for certain carefully constructed sequences. This slow mixing implies that, in the worst-case scenario, the decoder requires super-polynomial time to converge to the desired posterior distribution.
\end{abstract}

\section{Introduction}

Reed--Muller (RM) codes were invented in 1954 \cite{Muller54}, with the first decoding algorithm invented and published by Reed in the same year \cite{Reed54}. Being among the first codes ever constructed, RM codes have been the subject of extensive study in both electrical engineering and theoretical computer science, especially since the introduction of the closely-related, widely-adopted polar codes. Despite this, many questions about the performance of RM codes have only been answered recently. In 2017, Kudekar et al. showed that the Maximum a Posteriori (MAP) Decoding of RM codes achieves capacity on the binary erasure channel (BEC) \cite{Kudekar++2017}. Then, in 2021, Reeves and Pfister showed that, under MAP decoding, RM codes have vanishing bit-error probability on binary memoryless symmetric (BMS) channels \cite{Reeves--Pfister24}. Finally, in 2023, Sandon and Abbe extended the aforementioned result to show that RM codes have vanishing block-error probability on BMS channels, providing a positive answer to the question of whether RM codes achieve capacity on BMS channels \cite{Abbe--Sandon23}. The natural next step in this line of inquiry is to try to establish capacity-achieving performance for a polynomial-time RM decoder. This would help bridge the gap between the existing theory developed by \cite{Kudekar++2017, Reeves--Pfister24, Abbe--Sandon23} and practical use. Indeed, no existing polynomial-time decoders, such as Reed's original decoder, for RM codes are known to achieve the same block-error rate performance as MAP decoding for sequences of RM codes that achieve capacity over the BSC. For sequences of RM codes of fixed order, where the rate is vanishing and the number of codewords is polynomial in the blocklength--so that MAP decoding can be done in polynomial time, it is known that there are efficient algorithms, such as the Fast Hadamard Transform \cite{Green66} for order-1 RM codes, that achieve the same block-error rate performance as MAP decoding. A recent work of Rameshwar and Lalitha implies that, for RM codes of fixed order $r$, the Recursive Projection Aggregation (RPA) algorithm of Ye and Abbe \cite{Ye--Abbe19}, with complexity $O(n^r\log(n))$, also achieves the same block error-rate performance as the MAP decoder\cite{Rameshwar--Lalitha24}. 
Other practical decoders, such as
Dumer's recursive decoding algorithms \cite{Dumer04} and Sakkour's algorithm \cite{Sakkour05}, have focused on the short-to-medium blocklength regime and particular ranges of $r$, and either lack time-complexity guarantees in the asymptotic setting, or error-rate guarantees for longer blocklengths.

\subsection{Decoding by Posterior Sampling}
In this paper, we study a specific type of decoding algorithm based on posterior sampling.
To state our problem setting precisely, we consider the channel coding problem, where some message $\mathbf{m} \in \mathcal{M}$ is encoded as $\mathbf{x}(\mathbf{m}) \in \mathcal{X}^n$ and then transmitted over some discrete memoryless channel $p(y|x)$. Assuming a uniform prior on the set of messages and given a received sequence $\mathbf{y} \in \mathcal{Y}^n$, the MAP decoder produces an $\hat{\mathbf{m}} \in \mathcal{M}$ according to the rule $\hat{\mathbf{m}} = \arg\max_{\mathbf{m}' \in \mathcal{M}}p(\mathbf{x}(\mathbf{m})|\mathbf{y})$. This is optimal in the sense that the probability of error $P_e = \mathrm{Pr}[\mathbf m \ne \hat {\mathbf {m}}]$ is minimized. By the following lemma, 
however, it turns out that the asymptotic performance of MAP decoding can be matched by sampling from the posterior $p(\mathbf{m}|\mathbf{y})$, rather than selecting the message with the largest probability mass. This is formalized in the below statement.
\begin{lemma}[Lemma 3 in \cite{Kudekar16}, Theorem 7 in \cite{Liu--Cuff--Verdu2017}]
    Let $P_e^*$ be the optimal probability of error (ie. the one given by MAP decoding), and let $M' \sim p(\mathbf{m}|\mathbf{y})$. Then $$P^*_e \le \mathrm{Pr}[M' \ne \mathbf{m}] \le 2P^*_e.$$
\end{lemma}
Thus, if a sequence of codes (such as RM codes) is capacity-achieving under MAP decoding, then they are also capacity achieving under posterior sampling.

It still remains to sample from the posterior distribution, which is often computationally infeasible to do directly, since the distributions in question are often high-dimensional and very nonuniform. We can instead sample through Markov Chain Monte Carlo (MCMC) methods, which consider a Markov Chain whose stationary distribution is the desired posterior $p(\mathbf{m}|\mathbf{y})$. If such a Markov chain is ergodic, then it will converge to the desired posterior. The usual measure of time complexity for such a Markov chain is its mixing time, which is the number of steps required to ensure that the resulting distribution is within some $\epsilon$ Total Variation (TV) distance of the stationary distribution, regardless of the initial distribution. If we take enough steps in a Markov chain to get within some $\epsilon \rightarrow 0$ TV distance of the posterior, then by the Liu-Cuff-Verd\'u lemma, posterior sampling through MCMC methods is capacity achieving if the underlying code is capacity achieving.

Using MCMC methods to decode was first proposed by Neal in 2001 \cite{Neal01}. The proposed algorithm considered Gibbs sampling on the codeword space, with some proposed modifications to decrease the mixing time of the decoder, and to ensure that the resulting Markov Chain was indeed ergodic. A recent paper by Jain, Rameshwar, and Kashyap considered using MCMC methods, specifically an annealed Metropolis-Hastings chain on the codeword space, to estimate the weight enumerators of RM codes \cite{Jain--Rameshwar--Kashyap24}. We note that Algorithm 2 in their work can be appropriately modified to yield an MCMC decoder (where there is a nonzero transition probability between two codewords if and only if they differ by some minimum-weight nonzero codeword or the zero codeword) whose stationary distribution is the desired posterior.
Huang explored, in his PhD Thesis \cite{Huang2023}, many MCMC decoders in the short blocklength regime, based on running a Markov Chain on the message space. 
Many of these schemes are based on a basic MCMC decoder, known as the Gibbs decoder or Glauber dynamics, where a message bit is chosen uniformly at random, and then resampled according to the marginal posterior obtained by fixing the other bits to their assigned values. 
In this paper we provide an analysis of the mixing time of the Gibbs decoder. 

\subsection{Main Contribution}
Our results demonstrate that for specific, carefully constructed channel output sequences, the Gibbs decoder exhibits slow mixing. This implies that, in the worst-case scenario, the Gibbs decoder may struggle to converge in polynomial time to the desired posterior distribution when decoding RM codes over the binary symmetric channel.

\section{Preliminaries}

\subsection{Reed--Muller (RM) codes}
\label{sec:Reed-Muller}
We begin by establishing the notation for RM codes. RM codes belong to the family of polynomial codes. We write $\F_2[z_1, \ldots, z_m]$ for the ring of $m$-variate polynomials over the finite field $\F_2$.
Define the \emph{evaluation vector} of the polynomial $f\in \F_2[z_1, \ldots, z_m]$, $\operatorname{Eval}(f)$, as a binary vector whose coordinates corresponding to the evaluations of $f$ at all $2^{m}$ vectors in $\F_2^m$, usually in lexicographic ordering.
The \emph{RM code $\operatorname{RM}(r, m)$} is defined by a $k \times n$ generator matrix $G$, where each row of $G$ is the evaluation vector of a monomial $f \in \F_2[z_1, \ldots, z_m]$ of degree at most $r$.
Hence, the RM code $\mathrm{RM}(r, m)$ is explicitly
$$
\mathrm{RM}(r, m) = \left\{\operatorname{Eval}(f): f \in \mathbb{F}_2\left[z_1, \ldots, z_m\right], \operatorname{deg}(f) \leq r\right\}.
$$
For example, $\operatorname{RM}(2,3)$ is defined by the following generator matrix:
$$
G=\left[\begin{array}{c}
\operatorname{Eval}\left(z_1 z_2\right) \\
\operatorname{Eval}\left(z_1 z_3\right) \\
\operatorname{Eval}\left(z_2 z_3\right) \\
\operatorname{Eval}\left(z_1\right) \\
\operatorname{Eval}\left(z_2\right) \\
\operatorname{Eval}\left(z_3\right) \\
\operatorname{Eval}(1)
\end{array}\right]=\left[\begin{array}{llllllll}
1 & 1 & 0 & 0 & 0 & 0 & 0 & 0 \\
1 & 0 & 1 & 0 & 0 & 0 & 0 & 0 \\
1 & 0 & 0 & 0 & 1 & 0 & 0 & 0 \\
1 & 1 & 1 & 1 & 0 & 0 & 0 & 0 \\
1 & 1 & 0 & 0 & 1 & 1 & 0 & 0 \\
1 & 0 & 1 & 0 & 1 & 0 & 1 & 0 \\
1 & 1 & 1 & 1 & 1 & 1 & 1 & 1
\end{array}\right].
$$
Since $z^2 = z$ for all $z \in \F_2$ it suffices to consider polynomials whose terms have degree 1 in each indeterminate. Consequently, the dimension $k$ and length $n$ satisfy
$$
k=\binom{m}{\leq r}:=\sum_{i=0}^r\binom{m}{i} \text { and } n=2^m .
$$
The rate of the RM code $\operatorname{RM}(r, m)$ is given by $R = 2^{-m}\binom{m}{\leq r}$, and the minimum distance between distinct codewords is $2^{m - r}$.

Consider transmission over a Binary Symmetric Channel (BSC) with crossover probability $p$. Let $\theta = \frac{p}{1 - p}$ and assume that messages $\mathbf{m} \in \mathbb{F}_{2}^{k}$ are generated uniformly at random. Denote by $\mathbf{y} \in \mathbb{F}_{2}^{n}$ the received sequence and $d_{H}(\mathbf{x}, \mathbf{y})$ the Hamming distance between binary vectors $\mathbf{x}$ and $\mathbf{y}$. The posterior probability of a message $\mathbf{m}$ given the received sequence $\mathbf{y}$ is expressed as
\begin{align*}
    p(\mathbf{m} \mid \mathbf{y})& =\frac{p(\mathbf{y} \mid \mathbf{m}) p(\mathbf{m})}{p(\mathbf{y})} \\ & =\frac{p^{d_H(\mathbf{m} G, \mathbf{y})} \cdot(1-p)^{n-d_H(\mathbf{m} G, \mathbf{y})}}{2^k p(\mathbf{y})}, \\ &
    =\theta^{d_H(\mathbf{m} G, \mathbf{y})} \cdot \frac{(1-p)^n}{2^k p(\mathbf{y})}.
\end{align*}
This implies that the posterior probability of $\mathbf{m}$ given $\mathbf{y}$ is proportional to $\theta^{d_{H}(\mathbf{m}G, \mathbf{y})}$:
$$
p(\mathbf{m} \mid \mathbf{y}) \propto \theta^{d_H(\mathbf{m} G, \mathbf{y})}.
$$

\subsection{Markov Chain and Mixing Time}
\label{sec:MCMT}
Now, we introduce some basic properties of Markov Chains. Let $P$ denote the transition matrix of a Markov chain on a finite state space $\Omega$ and $P^t$ denote the $t$-fold matrix multiplication. Let $P(s_1 \rightarrow s_2)$ represent the transition probability from state $s_1$ to state $s_2$. We say $P$ is \emph{ergodic} if it satisfies the following conditions:
\begin{itemize}
    \item (Irreducibility) For all $s_1, s_2 \in \Omega$, there exists a non-negative integer $t$ such that $P^{t}(s_1 \rightarrow s_2) > 0$, and,
    \item (Aperiodicity) For all $s \in \Omega$, the greatest common divisor of the set $\{t \geq 1 \colon P^{t}(s \rightarrow s) > 0\}$ is equal to $1$.
\end{itemize}
A distribution $\mu$ on $\Omega$ is said to be \emph{stationary} with respect to $P$ if $\mu P = \mu$. 
The following theorem gives a relationship between ergodicity and the uniqueness of, and convergence to, a stationary state:
\begin{theorem} [Corollary 1.17, Theorem 4.9~\cite{levin2017markov}] 
\label{thm:fundthm}
    Let $P$ be an ergodic Markov chain on a finite state space $\Omega$. Then $P$ possesses a unique stationary distribution $\mu$ on $\Omega$, and for any initial distribution $\mu_0$, the distribution $\mu_t = \mu_0 P^t$ of the state $X_t$ converges pointwise to $\mu$ as $t \to \infty$.
\end{theorem}
Let $\epsilon>0$. The \emph{$\epsilon$-mixing time of $P$}
is defined as
$$
T_{\operatorname{mix}}(\epsilon):=\sup _{\mu_0} \min \left\{t \geq 0:\left\|\mu_0 P^t-\mu\right\|_{\mathrm{TV}} \leq \epsilon\right\},
$$
where the supremum is taken over all possible initial distributions $\mu_0$. Here, $\|\cdot\|_{\mathrm{TV}}$ represents the total variation distance between distributions.

We further define the \emph{total variation mixing time}, or \emph{mixing time} for short, of $P$ as
$$
T_{\operatorname{mix}}:=T_{\operatorname{mix}}\left(\tfrac{1}{4}\right).
$$
To analyze the slow mixing, it suffices to analyze the total variation mixing time, as this provides a lower bound for the mixing time for any $\epsilon \to 0$. 

Next we define the \emph{bottleneck ratio}, also known as \emph{conductance}. This quantity provides a means of bounding the mixing time of a Markov Chain from below and is instrumental in analyzing slow mixing phenomena. Let $P$ be an ergodic Markov chain with stationary distribution $\mu$ on a state space $\Omega$. The \emph{bottleneck ratio of a subset} $\mathcal{S} \subseteq \Omega$ is defined as
$$
\Phi(\mathcal{S}):=\frac{\sum_{s_1 \in \mathcal{S}, s_2 \in \Omega \backslash \mathcal{S}} \mu(s_1) P(s_1 \rightarrow s_2)}{\sum_{s_1 \in \mathcal{S}} \mu(s_1)},
$$
where $\mu$ denotes the stationary distribution of $P$. The \emph{bottleneck ratio of the Markov chain $P$} is then given by taking the infimum of the bottleneck ratio over all subsets of $\Omega$ with probability at most $\frac12$:
$$
\Phi(P):=\inf _{\substack{\mathcal{S} \in \Omega \\ \mu(S) \leq \frac{1}{2}}} \Phi(\mathcal{S}).
$$
We can use the bottleneck ratio of a Markov chain $P$ to bound its mixing time.
\begin{theorem}[Theorem 7.4~\cite{levin2017markov}]\label{7.4}
    The mixing time $T_{\operatorname{mix}}$ of $P$ satisfies
    $$
    T_{\operatorname{mix}} \geq \frac{1}{4 \Phi(P)} .
    $$
\end{theorem}

\subsection{Gibbs Decoding}
\label{sec:GD}
The Gibbs decoder is an MCMC method that recovers the message by sampling the message space $\mathbb{F}_{2}^{k}$ according to the posterior distribution $p(\mathbf{m}|\mathbf{y})$. For any message $\mathbf{m} \in \mathbb{F}_{2}^{k}$, write $\mathbf{m}^{\oplus i}$ for the message obtained by flipping the $i$-th coordinate of $\mathbf{m}$ (modulo $2$), while keeping all other coordinates unchanged. Let $\mu$ be the posterior distribution on $\F_2^k$, that is $\mu(\mathbf{m})=p(\mathbf{m} \mid \mathbf{y})$, where $\mathbf{y}$ is the received sequence. In each iteration, the Gibbs sampler performs the following:
\begin{enumerate}
    \item Select a coordinate $i \in [k]$ uniformly at random.
    \item Set $m_i$ to $m_i \oplus 1$ with probability $\frac{\mu(\mathbf{m}^{\oplus i})}{\mu(\mathbf{m}) + \mu(\mathbf{m}^{\oplus i})}$, and set $m_i$ to $m_i$ with remaining probability.
\end{enumerate}
Thus, the transition probabilities are given by
\begin{align*}
    P\left(\mathbf{m} \rightarrow \mathbf{m}^{\oplus i}\right)&=\frac{1}{k} \cdot \frac{\mu\left(\mathbf{m}^{\oplus i}\right)}{\mu(\mathbf{m})+\mu\left(\mathbf{m}^{\oplus i}\right)}, \quad \forall i \in[k],\\
    P(\mathbf{m} \rightarrow \mathbf{m})&=1-\sum_{i=1}^k P\left(\mathbf{m} \rightarrow \mathbf{m}^{\oplus i}\right),
\end{align*}
and for all the other $\mathbf{m}'$, we have $P(\mathbf{m}\rightarrow \mathbf{m}')=0$.
Since
\begin{align*}
    \mu(\mathbf{m}) \cdot P\left(\mathbf{m} \rightarrow \mathbf{m}^{\oplus i}\right)&=\frac{1}{k} \cdot \frac{\mu(\mathbf{m}) \cdot \mu\left(\mathbf{m}^{\oplus i}\right)}{\mu(\mathbf{m})+\mu\left(\mathbf{m}^{\oplus i}\right)}\\ &=\mu\left(\mathbf{m}^{\oplus i}\right) \cdot P\left(\mathbf{m}^{\oplus i} \rightarrow \mathbf{m}\right),
\end{align*}
we have $\mu P=\mu$, which implies that $\mu$ is a stationary distribution of $P$. To show the uniqueness of $\mu$ and convergence to $\mu$, it suffices to show, by Theorem \ref{thm:fundthm}, that $P$ is ergodic.
\begin{lemma}
    The Gibbs Sampler $P$ defined through the posterior $p(\mathbf{m} | \mathbf{y})$ above is ergodic, for any received sequence $\mathbf{y} \in \F_2^n$.
\end{lemma}
\begin{IEEEproof}
    First, note that for any $\mathbf{y} \in \F_2^n$, $\mu(\mathbf{m}) > 0$ for all $\mathbf{m} \in \F_2^k$, so that $0 < P(\mathbf{m} \rightarrow \mathbf{m}^{\oplus i}) < 1/k$ for all $\mathbf{m} \in \F_2^k$ and $i \in [k]$.
    To see that $P$ is aperiodic, note that $P(\mathbf{m} \rightarrow \mathbf{m}^{\oplus i}) < 1/k$ for all $i \in [k]$ and $\mathbf{m} \in \F_2^k$. Hence, $\sum_{i=1}^kP(\mathbf{m} \rightarrow \mathbf{m}^{\oplus i}) < 1$ and $P(\mathbf{m} \rightarrow \mathbf{m}) > 0$ for all $\mathbf{m} \in \F_2^k$. Thus, for each $\mathbf{m} \in \F_2^k$, $1 \in \set{t \ge 1 : P^t(\mathbf{m} \rightarrow \mathbf{m}) > 0}$ so that $\gcd\set{t \ge 1 : P^t(\mathbf{m} \rightarrow \mathbf{m}) > 0} = 1$. To see that $P$ is irreducible, note that $P(\mathbf{m} \rightarrow \mathbf{m}^{\oplus i}) > 0$ for all $i \in [k]$ and $\mathbf{m} \in \F_2^k$, so that for any $\mathbf{m}, \mathbf{m}' \in \F_2^k$, and any $j \ge d_H(\mathbf{m}, \mathbf{m}')$, $P^j(\mathbf{m} \rightarrow \mathbf{m}') > 0$.
\end{IEEEproof}

\section{Main Results}
Recall that the set of conditionally $\varepsilon$-typical length $n$ sequences for $\operatorname{BSC}(p)$ with input $\mathbf{x}$ is defined as 
$$
\mathcal{T}_\varepsilon^{(n)}(Y|\mathbf{x})=\left\{\mathbf{y}:np(1-\varepsilon)\leq d_{H}(\mathbf{x},\mathbf{y})\leq np(1+\varepsilon)\right\}.
$$
We prove the following theorem for the worst-case performance of the Gibbs Decoder.
\begin{theorem}
    \label{mainthm}
    Given any crossover probability $p < \frac{1}{2}$ and rate $R<1-H(p)$, for any sequence of RM codes $\mathrm{RM}\left(r_j, m_j\right)$ satisfying $\lim _{j \rightarrow \infty} m_j=\infty$ and $\lim _{j \rightarrow \infty} R_j=R$, there exists a sequence of pairs 
    $\left\{\left(\mathbf{m}_j G_j, \mathbf{y}_j\right)\right\}$ satisfying 
    \begin{enumerate}
        \item $\mathbf{y}_j\in \mathcal{T}_{1 / 2}^{(n_j)}(Y|\mathbf{m}_j G_j)$ and 
        \item the mixing time for running the Gibbs decoder on $\mathbf{y}_j$ using $\mathrm{RM}\left(r_j, m_j\right)$ is
        $$
        T_{\operatorname{mix}}=\Omega(\exp (\sqrt{n} \cdot \exp (-\sqrt{\log (n)}))).
        $$
    \end{enumerate}
\end{theorem}
\begin{remark}
    This result suggests that the worst-case decoding complexity is superpolynomial. Although the $\mathbf{y}$ we construct is a typical received sequence for some codeword, the probability of receiving such $\mathbf{y}$ might be vanishing. Thus, it may be the case that the average-case decoding complexity for the Gibbs decoder is polynomial. Numerical evidence, however, suggests that the average-case convergence rate of the Gibbs decoder is also slow \cite[Section~3.3.3]{Huang2023}.
\end{remark}
\begin{remark}
    We also note that there are several variations of the Gibbs decoder that show empirically faster convergence to the posterior. For instance, numerical simulations have shown that block updates (i.e. updating multiple bits in each iteration) decreases the number of iterations required to converge (although sampling from the conditional posterior for multiple bits is harder than for a single bit). Another method of improving the convergence is to consider an annealing schedule, so that at the $t$-th iteration, we consider the Gibbs update given by the distribution $p_{\alpha(t)}(\mathbf{m}|\mathbf{y}) \propto p(\mathbf{m}| \mathbf{y})^{\alpha(t)}$ for some $\alpha(t) \in [0, 1]$. These methods are covered in greater detail in \cite{Huang--Kim23}, where the authors run multiple annealed block Gibbs decoders on RM codes, on different subsets of $\F_2^k$ in parallel, to improve the mixing time. While our mixing time analysis seems to rule out initializing Gibbs decoding at an arbitrary distribution for RM codes, \cite{Huang--Kim20} considers using the Gibbs decoder after performing belief propagation on LDPC codes. As post-processing, this is able to overcome the error-floor phenomenon associated with belief propagation for some codes.
\end{remark}

\section{Proof of Theorem~\ref{mainthm}}
\label{sec:pf_main}
The proof of Theorem~\ref{mainthm} is divided into two key steps.
In the first step, Lemma~\ref{lem3} provides a sufficient condition for the Gibbs decoder to exhibit slow mixing. If there exists a message $\mathbf{m}$ with posterior probability $\mu(\mathbf{m}) \leq \frac{1}{2}$, and flipping any bit of $\mathbf{m}$ results in a significant increase in the Hamming distance to the received sequence $\mathbf{y}$, then the decoder struggles to move away from $\mathbf{m}$. The rationale is that messages around $\mathbf{m}$ have much smaller posterior probabilities, effectively ``trapping'' the decoder near $\mathbf{m}$. As a result, the decoder cannot explore the space of messages with significant posterior probability, leading to slow mixing.

In the second step, we explicitly construct such a typical received sequence $\mathbf{y}$ for any RM code $\mathrm{RM}(r, m)$ with sufficiently large $m$ in Lemma~\ref{lem4}. 
Specifically, when the current message is the all-zero message $\mathbf{0}$, we construct a received sequence $\mathbf{y}$ that ensures that flipping any bit of the all-zero message results in a significant increase in the Hamming distance to $\mathbf{y}$. 

\begin{lemma}
    \label{lem3}
    Consider the Reed--Muller code $\mathrm{RM}(r, m)$ with generator matrix $G$. Let $\mathbf{y} \in \mathbb{F}_2^n$. If there exists a message $\mathbf{m} \in \mathbb{F}_2^k$ such that $\mu(\mathbf{m}) \leq \frac{1}{2}$, and
    $$
    \delta_i(\mathbf{m}):=d_H\left(\mathbf{m}^{\oplus i} G, \mathbf{y}\right)-d_H(\mathbf{m} G, \mathbf{y})=\Omega(f(n)), \, \,\, \forall i \in [k],
    $$
    then the mixing time of the corresponding Gibbs decoder on $\mathbf{y}$ is $T_{\operatorname{mix}}=\Omega\left(e^{f(n)}\right)$.
\end{lemma}

\begin{lemma}
    \label{lem4}
    Given a crossover probability $p < \frac{1}{2}$, let $q=\left\lceil\log _2 \frac{1}{p}\right\rceil$. For the RM code $\mathrm{RM}(r, m)$ with $m \geq q$, there exists a non-zero vector $\mathbf{y} \in \mathbb{F}_2^m$ such that:
    \begin{enumerate}
        \item There exists a codeword $\mathbf{c}$ such that $d_H(\mathbf{c}, \mathbf{y})=2^{m-q}$;
        \item The posterior distribution for the all-zero message given received sequence $\mathbf{y}$ satisfies $\mu(\mathbf{0}) \leq 0.5$; and
        \item Adding any row $\mathbf{g}_i$ of the generator matrix\footnote{We consider the standard generator matrix where each row corresponds to a monomial of degree at most $r$ as introduced in Section~\ref{sec:Reed-Muller}.} of $\mathrm{RM}(r, m)$ to $\mathbf{y}$ increases its weight by at least $2^{m-r-q+1}$, i.e.,
        $$
        \operatorname{wt}(\mathbf{y}+\mathbf{g}_i)-\operatorname{wt}(\mathbf{y})\geq \frac{d}{2^{q-1}}=2^{m-r-q+1}, \quad \forall i\in[k].
        $$
    \end{enumerate}
\end{lemma}

The key idea in establishing Lemma~\ref{lem3} is to identify a singleton message set $\mathcal{S}$, whose bottleneck ratio gives a lower bound of the bottleneck ratio of $P$. 

\begin{IEEEproof}[{\bf Proof of Lemma 3}]
    For a subset of messages $\mathcal{S} \subseteq \mathbb{F}_2^k$ with $\mu(\mathcal{S}) \leq \frac{1}{2}$, the bottleneck ratio can be expressed as
    $$
    \Phi(\mathcal{S})=\frac{\sum_{\mathbf{m} \in \mathcal{S}} \mu(\mathbf{m})[1-P(\mathbf{m} \rightarrow \mathbf{m})]}{\mu(\mathcal{S})}.
    $$
    Consider the special case where $\mathcal{S}=\{\mathbf{m}\}$ is a singleton set containing only one state $\mathbf{m}$, such that $\mu(\mathbf{m}) \leq \frac{1}{2}$. In this scenario, the bottleneck ratio simplifies to
    $$
    \Phi(\mathcal{S})=1-P(\mathbf{m} \rightarrow \mathbf{m}).
    $$
    Define $\delta_i(\mathbf{m}):=d_H\left(\mathbf{m}^{\oplus i} G, \mathbf{y}\right)-d_H(\mathbf{m} G, \mathbf{y})$, where $d_H$ denotes the Hamming distance. Then, the probability $1-P(\mathbf{m} \rightarrow \mathbf{m})$ can be written as
    \begin{align*}
        1-P(\mathbf{m} \rightarrow \mathbf{m}) & =\frac{1}{k} \sum_{i=1}^k\frac{\mu\left(\mathbf{m}^{\oplus i}\right)}{\mu(\mathbf{m})+\mu\left(\mathbf{m}^{\oplus i}\right)}  
        =\frac{1}{k} \sum_{i=1}^k \tfrac{\tfrac{\mu\left(\mathbf{m}^{\oplus i}\right)}{\mu(\mathbf{m})}}{1+\tfrac{\mu\left(\mathbf{m}^{\oplus i}\right)}{\mu(\mathbf{m})}} .
    \end{align*}
    Using the inequality $\frac{x}{1+x} \leq x$ for $x \geq 0$, we obtain
    $$
    1-P(\mathbf{m} \rightarrow \mathbf{m}) \leq \frac{1}{k} \sum_{i=1}^k \frac{\mu\left(\mathbf{m}^{\oplus i}\right)}{\mu(\mathbf{m})}.
    $$
    From the proportionality relation on the posterior, we may write the ratio of posteriors as $\theta^{\delta_i(\mathbf{m})}=\frac{\mu\left(\mathbf{m}^{\oplus i}\right)}{\mu(\mathbf{m})}$, where $\theta=\frac{p}{1-p} < 1$ is a parameter given by the channel. Substituting this, we get
    $$
    1-P(\mathbf{m} \rightarrow \mathbf{m}) \leq \frac{1}{k} \sum_{i=1}^k \theta^{\delta_i(\mathbf{m})}.
    $$
    From the lower bound Theorem~\ref{7.4} on mixing time, we know
    $$
    T_{\operatorname{mix}} \geq \frac{1}{4 \Phi(P)}=\frac{1}{4(1-P(\mathbf{m} \rightarrow \mathbf{m}))}.
    $$
    Therefore, if $\delta_i(\mathbf{m})=\Omega(f(n))$ for all $i$, then
    $$
    1-P(\mathbf{m} \rightarrow \mathbf{m})=O(\exp (-f(n)))
    $$
    and consequently
    $
    T_{\operatorname{mix}}=\Omega(\exp (f(n))).
    $
\end{IEEEproof}

\begin{IEEEproof}[{\bf Proof of Lemma 4}]
    Recall that in an RM Code, each codeword is the evaluation vector of an $m$-variate polynomial $f \in \mathbb{F}_2\left[z_1, z_2, \ldots, z_m\right]$ of total degree at most $r$ over all $2^m$ vectors $\mathbf{z}=\left(z_1, z_2, \ldots, z_m\right) \in \mathbb{F}_2^m$.
    Define a $m$-variate polynomial 
    \begin{equation}
    \textstyle{
    f_{\mathbf{y}}\left(z_1, \ldots, z_m\right)=\left(\prod_{i=1}^{q-1} z_i+1\right)\left(z_m+1\right).}
    \end{equation}
    Let $\mathbf{y}=\operatorname{Eval}(f_{\mathbf{y}})$ denote the evaluation vector of $f_{\mathbf{y}}$.  Since $f_{\mathbf{y}}(\mathbf{0})=1$, it follows that $\mathbf{y} \neq \mathbf{0}$.

    Next, consider the evaluation vector of the $m$-variate polynomial $z_m+1$, denoted as $\mathbf{c}\triangleq \operatorname{Eval}(z_m+1)$. Since this is a degree-1 polynomial, its evaluation vector $\mathbf{c}$ is a codeword for any RM code $\mathrm{RM}(r, m)$ with $r\ge 1$. Let $\mathbf{u}$ be the corresponding message, i.e., $\mathbf{c} = \mathbf{u}G$. The vector $\mathbf{c}$ differs from $\mathbf{y}$ only at the coordinates $\mathbf{z}$ with $z_1=z_2=\cdots=z_{q-1}=1$ and $z_m=0$. With $q$ variables fixed, there are $2^{m-q}$ such coordinates. Therefore, the Hamming distance $d_H(\mathbf{c},\mathbf{y})=2^{m-q}$, establishing statement 1) in Lemma~\ref{lem4}.

    To establish statement 2), we analyze the posterior probabilities of the all-zero message and message $\mathbf{u}$.
    The posterior probability of the all-zero message $\mathbf{0}$ is proportional to $\theta^{\mathrm{wt}(\mathbf{y})}$. To compute $\mathrm{wt}(\mathbf{y})$, note that
    $f_\mathbf{y}(\mathbf{z})$ is 1 only when $z_m = 0$ and $\prod_{i=1}^{q-1}z_i = 0$, which implies $\mathrm{wt}(\mathbf{y}) = 2^{m-1} - 2^{m-q}$. Thus, the posterior probability of the all-zero message is proportional to $\theta^{2^{m-1}-2^{m-q}}$.
    Since $d_H(\mathbf{c},\mathbf{y})=2^{m-q}$, the posterior probability of message 
    $\mathbf{u}$ is proportional to $\theta^{2^{m-q}}$. For $p < \frac{1}{2}$, we have $q \geq 2$, which leads to $\theta^{2^{m-1}-2^{m-q}} \leq \theta^{2^{m-q}}$. This implies that the posterior probability $\mu(\mathbf{u})\ge\mu(\mathbf{0})$. If $\mu(\mathbf{0})>\frac{1}{2}$, it would result in a contradiction, as the posterior probabilities of both messages would sum to more than 1. Therefore, we conclude that $\mu(\mathbf{0}) \leq \frac{1}{2}$, establishing statement 2).

    Finally, to prove statement 3), we define $\mathcal{S}(f):=\left\{\mathbf{z} \in \mathbb{F}_2^m: f(\mathbf{z})=1\right\}$. Consider an arbitrary row $\mathbf{g}_i$ of $G$, which corresponds to the evaluation vector of some monomial $f \in \mathbb{F}_2\left[z_1, z_2, \ldots, z_m\right]$ with $\operatorname{deg}(f) \leq r$. The weight of $\mathbf{g}_i$ is equal to $|\mathcal{S}(f)|$. Denote $f=z_{j_1}z_{j_2}\cdots z_{j_t}$ where $\left\{j_1, \ldots, j_t\right\}$ is some subset of $[m]$ with size $t \le r$.
    There are two cases.

    Case 1: $m \in\left\{j_1, \ldots, j_t\right\}$. 
    In this case, for any evaluation coordinate $\mathbf{z} \in \mathbb{F}_2^m$ such that $f(\mathbf{z}) = 1$, we must have $z_m = 1$. This forces $f_\mathbf{y}(\mathbf{z}) = 0$ since $f_\mathbf{y} =(\prod_{i=1}^{q-1}z_i + 1)(z_m+1)$.
    Thus, $\mathcal{S}(f) \cap \mathcal{S}\left(f_{\mathbf{y}}\right)=$ $\emptyset$ and we have
    \begin{align*}
        \operatorname{wt}(\mathbf{y}+\mathbf{g}_i)-\operatorname{wt}(\mathbf{y})& =
        \left|\mathcal{S}\left(f+f_{\mathbf{y}}\right)\right|-\left|\mathcal{S}\left(f_{\mathbf{y}}\right)\right|\\
        & =|\mathcal{S}(f)|-2\left|\mathcal{S}(f) \cap \mathcal{S}\left(f_{\mathbf{y}}\right)\right| \\
        &=|\mathcal{S}(f)| \geq 2^{m-t} \geq 2^{m-r}.
    \end{align*}

    Case 2: $m \notin\left\{j_1, \ldots, j_t\right\}$. Let $\mathcal{I}:=\left\{j_1, \ldots, j_t\right\} \cap\{1,2, \ldots, q-1\}$. We have $0 \leq|\mathcal{I}| \leq \min \{t, q-1\}$. Then,
    \begin{align*}
    \mathcal{S}(f) \cap \mathcal{S}\left(f_{\mathbf{y}}\right)=\{& \mathbf{z} \in \mathbb{F}_2^m:  z_{j_1}=\cdots=z_{j_t}=1, z_m=0 , \\
     & \text { and at least one of } z_1, \ldots, z_{q-1} \text { is } 0 \}.
    \end{align*}
Partition the indices into three disjoint sets:
    \begin{align*}
    \mathcal{A}=\left\{j_1, \ldots, j_t, m\right\},\;
    \mathcal{B}=[q-1] \backslash \mathcal{I},\; 
    \mathcal{C}=[m] \backslash(\mathcal{A} \cup \mathcal{B}),
    \end{align*}
    where we write $[\ell]$ for the set $\set{1,\ldots, \ell}$. The number of $\mathbf{z} \in \mathbb{F}_2^m$ where all variables in $\mathcal{A}$ are fixed and at least one variable in $\mathcal{B}$ is 0 is:
    \begin{align*}
    \left|\mathcal{S}(f) \cap \mathcal{S}\left(f_{\mathbf{y}}\right)\right|&=(1)\left(2^{q-1-|\mathcal{I}|}-1\right)\left(2^{m-t-q+|\mathcal{I}|}\right)\\
    &=2^{m-t-1}-2^{m-t-q+|\mathcal{I}|} .
    \end{align*}
    Thus, we have
    \begin{align*}
    \operatorname{wt}(\mathbf{y}+\mathbf{g}_i)-\operatorname{wt}(\mathbf{y})& =\left|\mathcal{S}\left(f+f_{\mathbf{y}}\right)\right|-\left|\mathcal{S}\left(f_{\mathbf{y}}\right)\right|\\ & =|\mathcal{S}(f)|-2\left|\mathcal{S}(f) \cap \mathcal{S}\left(f_{\mathbf{y}}\right)\right| \\
    & =2^{m-t}-2\left(2^{m-t-1}-2^{m-t-q+|\mathcal{I}|}\right) \\
    & =2^{m-t-(q-1-|\mathcal{I}|)}  \geq 2^{m-r-q+1}.\\[-3em]
    \end{align*}
\end{IEEEproof}

With the two lemmas established, the final step of the proof involves analyzing the asymptotic expression of $2^{m-r-q+1}$ in terms of the codeword length $n=2^m$.

\begin{IEEEproof}[{\bf Proof of Theorem 3}]
    Let $p < \frac{1}{2}$ and $q=\left\lceil\log _2 \frac{1}{p}\right\rceil$. For any $m \geq q$ and the corresponding RM code $\mathrm{RM}(r, m)$, use Lemma~\ref{lem4} to construct $\mathbf{y} \neq 0$. There exists a codeword $\mathbf{c} \in \mathbb{F}_2^n$ such that $d_H(\mathbf{c}, \mathbf{y})=2^{m-q}$. Moreover, since
    $$
    \frac{1}{2}p\leq 2^{-\lceil \log_{2}(1/p) \rceil }\leq \frac{3}{2}p,
    $$
    there exists a message $\mathbf{m}$ such that $\mathbf{m} G=\mathbf{c}$ and $(\mathbf{m} G, \mathbf{y}) \in \mathcal{T}_{1 / 2}^{(n)}(Y|\mathbf{m} G)$. 
    
    For all $i \in[k]$, consider the $i$-th row $\mathbf{g}_i$ of $G$. From Lemma~\ref{lem4},
    $$
    \mathrm{wt}\left(\mathbf{y}+\mathbf{g}_i\right)-\mathrm{wt}(\mathbf{y}) \geq 2^{m-r-q+1} .
    $$
    This implies $\delta_i \geq 2^{m-r-q+1}$, where $\mathbf{m}_0=\mathbf{0} \in \mathbb{F}_2^k$.
    
    To analyze the asymptotic behavior of $2^{m-r-q+1}$, note that the binomial distribution satisfies:
    $$
    \operatorname{Pr}[\operatorname{Bin}(m, 0.5)=r]=\frac{\binom{m}{r}}{2^m}.
    $$
    Let $k=\binom{m}{\leq r}$. For large $m$, the rate converges to $R$, i.e., $\frac{k}{2^m} \rightarrow R$. Approximating via the normal distribution:
    \begin{align} \label{normal}
        \operatorname{Pr}\left[\mathcal{N}\left(\frac{m}{2}, \frac{m}{4}\right) \leq r\right]=\Phi\left(\frac{r-\frac{m}{2}}{\frac{\sqrt{m}}{2}}\right)=R,
    \end{align}
    where $\Phi$ is the cumulative distribution function of the standard normal distribution. By applying the inverse of $\Phi$ to both sides of equation~\eqref{normal} and letting $c := \Phi^{-1}(R)$, we obtain:
    $$
    r = \frac{m}{2}+c \cdot \frac{\sqrt{m}}{2}.
    $$
    Thus,
    $
    \delta_i \geq 2^{m-r-q} = 2^{(m-c \cdot \sqrt{m}) / 2-q}.
    $
    Since $n=2^m$, it follows from Lemma~\ref{lem3} that:
    $$
    T_{\operatorname{mix}}=\Omega(\exp (\sqrt{n} \cdot \exp (-\sqrt{\log (n)}))).
    $$
\end{IEEEproof}

\section{Conclusion}
In this paper, we have analyzed the worst-case complexity of the Gibbs decoder on RM codes. We have shown that, in the case of certain conditionally typical received sequences, Gibbs decoding has superpolynomial mixing time. For the posterior given by these received sequences, all-zeros message presents a bottleneck, preventing the Gibbs decoder from exploring the message space. Despite this, the question of whether other MCMC decoders can have polynomial-time complexity is left open.


\section*{Acknowledgment}
The authors are grateful to Wei Yu for encouraging them to work on this problem and Ryan Song for helpful discussions.




\bibliographystyle{IEEEtran}
\bibliography{references}

\end{document}